\documentclass{amsart}
\usepackage{amsmath}
\usepackage{amssymb,amsthm}
\usepackage{graphicx}
\usepackage{tikz}
\usepackage {threeparttable}
\usepackage[superscript]{cite}

\numberwithin{equation}{section}

\title[calculations of multi-center integrals]{Methods for accurate calculations of multi-center integrals of the squared Coulomb potential for lower bounds to energy levels of molecular systems}
\author{Sohei Ashida}
\begin{document}
\maketitle

\begin{abstract}
In this paper methods for calculations of  multi-center integrals of squared Coulomb potentials and Slater-type orbitals (STO) are derived. These integrals are necessary for accurate lower bounds to energy levels of molecular systems.  All multi-center integrals are reduced to fundamental integrals using the Gaunt coefficients and translation of STO. When the potential is the usual Coulomb potential, using the Laplace expansion or the Neumann expansion of the potential the integrals can be calculated. However, for the squared Coulomb potentials such expansions are not known. For the fundamental one-center and two-center integrals with squared Coulomb potentials, by methods free from such expansions exact analytic expressions and expressions by one-dimensional integrals of analytic functions are derived. The methods mainly rely on the integration in ellipsoidal coordinates, the Fourier transform, Hobson's theorem and expansion of differential operators by simple ones suitable for the calculation. Numerical results by these expressions are given and compared.
\end{abstract}

\section{Introduction}
As is well known, under the Born-Oppenheimer approximation properties of molecules such as molecular structures and rates of chemical reactions are understood from the dependence of electronic energy levels (i.e. eigenvalues of electronic Hamiltonians) on nuclear positions. Thus estimates for the eigenvalues of the Hamiltonian are central to studies of molecules. However, unless the eigenvalue problem is solved exactly, it is very difficult to evaluate the difference between the true eigenvalue and the estimates.

A method to obtain error estimates is to give both upper and lower bounds of the eigenvalues. In this method the true eigenvalue evidently lies between the two values. Upper bounds can be obtained by the variational method. Compared to upper bounds accurate lower bounds are much more difficult to achieve in many respects. Therefore, most of the results  obtained so far are concerned with variational upper bound, some perturbation theory or expansion theory whose error estimate is hopeless, or some approximation for bulk from a macroscopic viewpoint which is irrelevant to usual molecules.

However, only by the variatioal method reliable evaluation is impossible. Convergence itself is obvious if we use a complete system in a certain appropriate Hilbert space as a basis set, but it is the rate of the convergence that is important in practical calculations. One should note that seeming convergence of the value as the basis set increases does not necessarily imply the convergence to the true energy level, since addition of functions that does not contribute to the true eigenfunction to the basis set does not improve the value. In particular, there is no mathematically rigorous evidence that accurate upper bounds are obtained effectively using some basis set such as the Slater or Gaussian type orbitals ordered in a natural way or their linear combination obtained in some way,  and increasing the basis set.

Unfortunately, comparison of the upper bounds with experimental energy levels is also impossible except for equilibrium positions of the nuclei, because energy levels for unstable nuclear positions are difficult to determine experimentally accurately. In fact one of the most common purposes of calculations of energy levels is to predict the equilibrium geometry of molecules which is the minimum point of the sum of the energy level and the nuclear repulsion potential as a function of nuclear positions. For such a purpose we need a method which guarantees accuracy of the evaluation without resort to experimental data. Thus there should be a method of eigenvalue evaluation for which it is confirmed that the error from the true eigenvalue is very small in a mathematically rigorous way at least for small molecules. Therefore, methods for lower bounds are desirable.

In lower bound methods, Temple's inequality \cite{Te,Ka} is known to have high accuracy at least for simple systems. However, in order to apply Temple's inequality we need a lower bound of the eigenvalue next to the evaluated one. Thus we need to seek rough lower bounds by other methods in order to apply Temple's inequality. The most promising method for such lower bounds would be the Weinstein-Aronszajn intermediate problem method \cite{WS,A} or rather methods derived from that method \cite{BG,We} in which the Coulomb repulsion potentials between electrons are regarded as perturbation by a positive operator.

In these methods  (including the method by Temple's inequality), one needs to calculate the integral $\langle\Psi|H^2|\Psi\rangle$, where $\Psi$ is the wave function for $N$ electrons and $H$ is the Hamiltonian of all the electrons written in atomic units as
$$H=-\frac{1}{2}\sum_{i=1}^N\nabla_i^2-\sum_{i=1}^N\sum_{A=1}^M\frac{Z_A}{|\mathbf r_i-\mathbf R_A|}+\sum_{1\leq i<j\leq N}\frac{1}{|\mathbf r_i-\mathbf r_j|}.$$
Here $M$ is the number of nuclei, $\mathbf r_i$ and $\mathbf R_A$ are positions of electron $i$ and nucleus $A$ respectively, and $Z_A$ is the atomic number of nucleus $A$. The problem of evaluation of such integrals has been one of  the main difficulties in lower bound estimates \cite{Go, Ki} and has not been solved essentially so far.

As the function $\Psi$ some approximate eigenfunction of the Hamiltonian is used. Let $\mathbf r\in \mathbb R^3$ be a position of an electron and $\tilde {\mathbf r}\in \mathbb R^{3(N-1)}$ be the position of the other electrons. Kato \cite{Ka2} proved that a true eigenfunction $\Psi(\mathbf r,\tilde {\mathbf r})$ satisfies $\frac{\partial \hat\Psi^A}{\partial r_A}\Big|_{r_A=0}=-Z_A\Psi(\mathbf R_A,\tilde {\mathbf r})$ except at some points $\tilde {\mathbf r}$ of a set of lower dimension. Here $r_A=|\mathbf r-\mathbf R_A|$ and $\hat \Psi^A$ is the average value of $\Psi$ taken over the sphere $r_A=\mathrm{const}$ for a fixed value of $\tilde {\mathbf r}$. This well-known result called Kato's cusp condition implies that the true eigenfunction has cusps at the positions of nuclei like the eigenfunction of the hydrogen atom. Hence, for accurate evaluation of the energy level the Slater-type orbital (STO) which has a factor as $e^{-\zeta r}$ is suitable. Nevertheless, in practical calculations Gaussian-type orbitals (GTO) which has a factor as $e^{-\zeta r^2}$ are often used because the calculation of integrals for GTO is easier than that for STO.

In order to approximate the expectation value of squared Hamiltonian $H^2$ with respect to a Slater determinant of STO $\psi$ by that of a linear combination $\sum_i \phi_i$ of GTO we need to make the value of $\lVert (1-\nabla^2)(\psi-\sum_i\phi_i)\rVert\sim\lVert (1+k^2)(\mathcal F\psi-\sum_i\mathcal F\phi_i)\rVert$ small, where $\lVert u\rVert=\sqrt{\langle u,u\rangle}$ and $\mathcal F$ is the Fourier transform. However, the Fourier transform of STO is a rational function (cf. Eq. \eqref{myeq2.4.1}) in contrast to that the Fourier transform of GTO is again GTO. Thus for highly accurate estimate of the energy levels by GTO one clearly needs a huge basis set. Therefore, if all calculations can be accomplished efficiently for STO, there is no reason to use GTO.

The results in this paper are concerned with the evaluation of $\langle \Psi|H^2|\Psi\rangle$ where $\Psi$ is a linear combination of the Slater determinants $(N!)^{-1/2}\mathrm{det}\, |\psi_1\psi_2\dotsm \psi_N|$ of STO.  Here each $\psi_i$ is STO centered at one of the positions $\mathbf R_A,\ A=1,\dots, M$ of the nuclei. One of the most difficult integrals in the terms of $\langle \Psi|H^2|\Psi\rangle$ would be the integral of the following form:
\begin{equation}\label{myeq1.1}
\begin{split}
&[\psi_1(\mathbf r_A)\psi_2(\mathbf r_B)|\psi_3(\mathbf r'_C)\psi_4(\mathbf r'_D)]\\
&\quad=\int_{\mathbb R^3}\int_{\mathbb R^3}\psi_1^*(\mathbf r_A)\psi_2(\mathbf r_B)\frac{1}{|\mathbf r-\mathbf r'|^2}\psi_3^*(\mathbf r'_C)\psi_4(\mathbf r'_D)d\mathbf rd\mathbf r',
\end{split}
\end{equation}
where $\mathbf r_A:=\mathbf r-\mathbf R_A$ and $\mathbf r'_C:=\mathbf r'-\mathbf R_C$. If the squared Coulomb potential $\frac{1}{|\mathbf r-\mathbf r'|^2}$ is replaced by the usual Coulomb potential $\frac{1}{|\mathbf r-\mathbf r'|}$, this integral is the multi-center integral encountered in the calculations of variational upper bounds of molecular energy levels and is a central subject in the variational method.
A product of two STOs centered at the same point can be expressed as a finite sum of STOs using Gaunt coefficients \cite{Wen}, and STO centered at  $\mathbf R_A$ can be expanded by STOs centered at different point $\mathbf R_B$ \cite{RL, FS}. Thus the calculation of integral Eq. \eqref{myeq1.1} is reduced to that of the integrals of the following form:
\begin{equation}\label{myeq1.2}
[\psi_1(\mathbf r_A)|\psi_2(\mathbf r'_B)]=\int_{\mathbb R^3}\int_{\mathbb R^3}\psi_1^*(\mathbf r_A)\frac{1}{|\mathbf r-\mathbf r'|^2}\psi_2(\mathbf r'_B)d\mathbf rd\mathbf r'.
\end{equation}
In this paper analytic expressions and expressions by one-dimensional integrals of analytic functions are derived for this fundamental integral for both the cases of $\mathbf R_A=\mathbf R_B$ and $\mathbf R_A\neq \mathbf R_B$ (In this paper, the term "analytic expression" means an expression by functions for which efficient accurate evaluation have been well-established). If the factor $\frac{1}{|\mathbf r-\mathbf r'|^2}$ is replaced by the usual Coulomb potential $\frac{1}{|\mathbf r-\mathbf r'|}$, integral Eq. \eqref{myeq1.2} with $\mathbf R_A=\mathbf R_B$ can be calculated using the Laplace expansion \cite{Ja}
$$\frac{1}{|\mathbf r-\mathbf r'|}=4\pi\sum_{l=0}^{\infty}\sum_{m=-l}^l\frac{1}{2l+1}\frac{r_<^l}{r_>^{l+1}}Y^*_{lm}(\theta',\varphi')Y_{lm}(\theta,\varphi),$$
where $(r,\theta,\varphi)$ and $(r',\theta',\varphi')$ are polar coordinates of $\mathbf r$ and $\mathbf r'$ respectively, $r_<=\min\{r,r'\}$, $r_>=\max\{r,r'\}$, and $Y_{lm}$ is the spherical harmonics. Moreover, in this case for $\mathbf R_A=\mathbf R_C$, $\mathbf R_B=\mathbf R_D$ and $\mathbf R_A\neq \mathbf R_B$, integral Eq. \eqref{myeq1.1} can be calculated using the Neumann expansion \cite{Ne, Su, Ru}
\begin{align*}
\frac{1}{|\mathbf r-\mathbf r'|}&=\frac{2}{R}\sum_{l=0}^{\infty}\sum_{m=-l}^l(-1)^m(2l+1)\left(\frac{(l-|m|)!}{(l+|m|)!}\right)^2\\
&\quad\times P_l^{|m|}(\xi_<)Q_l^{|m|}(\xi_>)P_l^{|m|}(\eta) P_l^{|m|}(\eta')e^{im\varphi}e^{-im\varphi'},
\end{align*}
where $(\xi,\eta,\varphi)$ and $(\xi',\eta',\varphi')$ are ellipsoidal coordinates of $\mathbf r$ and $\mathbf r'$ respectively with foci $\mathbf R_A$ and $\mathbf R_B$, $\xi_<=\min\{\xi,\xi'\}$,  $\xi_>=\max\{\xi,\xi'\}$, $P_l^{|m|}$ and $Q_l^{|m|}$ are the associated Legeandre functions, and $R=|\mathbf R_A-\mathbf R_B|$. However, for the squared Coulomb potential such expansions have not been known. There has not been any substantial progress for the method to evaluate integral Eq. \eqref{myeq1.1} so far as far as the author knows. In fact it seems that such a study has long been abandoned due to its difficulty. In this paper we derive expressions for integral Eq. \eqref{myeq1.2} by methods free from such expansions. 

For both $\mathbf R_A=\mathbf R_B$ (one-center integral) and $\mathbf R_A\neq\mathbf R_B$ (two-center integrals) analytic expressions (for special cases of STO in the case of $\mathbf R_A=\mathbf R_B$) and expressions by one-dimensional integrals are derived. For both one-center and two-center integrals the analytic expressions are applicable only if the scaling parameters $\zeta$ of $\psi_1$ and $\psi_2$ are different. For the derivation of the expressions wide range of techniques are needed. In particular, we need integration in ellipsoidal coordinates, techniques concerning the Fourier transform, Hobson's theorem and expansion of differential operators by simple ones suitable for the calculation.

Numerical results by the analytic expression and the expression by one-dimensional integrals are compared. Accuracy of the expressions by one-dimensional integrals is much better than the analytic expressions  because of cancellation of significant digits in the analytic expressions. The accuracy would be reasonable for application to lower bound calculations of energy levels of small molecules.

\section{Definitions and basic formulas}
We consider the functions known as Slater type orbitals (STO). Let us denote the Cartesian coordinates of $\mathbf r\in \mathbb R^3$ by $x,y,z$. We also denote the polar coordinates of $\mathbf r$ by $r, \theta, \varphi$.  An unnormalized STO considered in this paper is defined by
$$\chi^n_{lm}(\mathbf r,\zeta)=Z_l^m(\mathbf r)r^{n-1}e^{-\zeta r},$$
where $n, l\in \mathbb N$, $m\in \mathbb Z$, $-l\leq m\leq l$, $\zeta>0$ is a parameter, and $Z_l^m(\mathbf r)$ is the spherical function defined by
$$Z^m_l(\mathbf r)=i^{m+|m|}r^lP_l^{|m|}(\cos \theta)e^{im\varphi},$$
that are homogeneous polynomials of $x,y,z$ of degree $l$. Here $P_l^m(t)$ is the associated Legendre function defined by $P_l^m(t)=(1-t^2)^{m/2}\frac{d^m}{dt^m}P_l(t)$, where $P_l(t)$ is the Legendre polynomial. It is well known that $Z_l^m$ satisfies the Laplace equation $\nabla^2 Z_l^m=0$. We also define $Y_{lm}(\theta,\varphi)$ by
\begin{align*}
Y_{lm}(\theta,\varphi)&=i^{m+|m|}\left(\frac{(2l+1)(l-|m|)!}{4\pi(l+|m|)!}\right)^{1/2}P_l^{|m|}(\cos \theta)e^{im\varphi}\\
&=\left(\frac{(2l+1)(l-|m|)!}{4\pi(l+|m|)!}\right)^{1/2}r^{-l}Z^m_l(\mathbf r).
\end{align*}
Then $Y_{lm}$ are spherical harmonics, and they are orthogonal to each other in $L^2(\mathbb S^2)$, i.e.
\begin{equation}\label{myeq2.0.1}
\int_0^{\pi}\int_0^{2\pi}Y_{lm}^*(\theta,\varphi)Y_{l'm'}(\theta,\varphi)\sin \theta d\theta d\varphi=\delta_{ll'}\delta_{mm'}.
\end{equation}

\section{Fundamental one-center integrals}\label{thirdsec}
The fundamental one-center integral $[\chi_{lm}^n|\chi_{l'm'}^{n'}]$ is defined by
$$[\chi_{lm}^n|\chi_{l'm'}^{n'}]=\int_{\mathbb R^3}\int_{\mathbb R^3}\frac{1}{|\mathbf r-\mathbf r'|^2}\overline{\chi_{lm}^n}(\mathbf r,\zeta)\chi_{l'm'}^{n'}(\mathbf r',\zeta')d\mathbf rd\mathbf r',$$
where $\overline{\chi_{lm}^n}$ is the complex conjugate of $\chi_{lm}^n$.
As we will see in Sect. \ref{sec2.2} this integral is not zero only if $l=l'$ and $m=m'$.

\subsection{Method 1: analytic expression for $l=m=0$}
For the fundamental one-center integrals with $l=m=0$ we have the following analytic expression.
\begin{equation}\label{myeq2.0.0.0.1}
[\chi_{0 0}^n|\chi_{0 0}^{n'}]=16\pi^2\left(-\frac{\partial}{\partial \zeta}\right)^n\left(-\frac{\partial}{\partial \zeta'}\right)^{n'}\frac{\log \zeta-\log \zeta'}{\zeta^2-\zeta'^2}.
\end{equation}
The formula Eq. \eqref{myeq2.0.0.0.1} is derived as follows.
Since $\chi_{00}^n=\left(-\frac{\partial}{\partial \zeta}\right)^n\chi_{0 0}^0$ we have only to prove
$$[\chi_{0 0}^0|\chi_{0 0}^0]=16\pi^2\frac{\log \zeta-\log \zeta'}{\zeta^2-\zeta'^2}.$$
By the change of variables $\tilde {\mathbf r}= -\mathbf r$, $\tilde {\mathbf r}'=\mathbf r'-\mathbf r$, we have
\begin{align*}
[\chi_{0 0}^0|\chi_{0 0}^0]&=\int_{\mathbb R^3}\int_{\mathbb R^3}\frac{1}{|\mathbf r-\mathbf r'|^2}\frac{e^{-\zeta |\mathbf r|}}{|\mathbf r|}\frac{e^{-\zeta' |\mathbf r'|}}{|\mathbf r'|}d\mathbf rd\mathbf r'\\
&=\int_{\mathbb R^3}\int_{\mathbb R^3}\frac{1}{|\tilde {\mathbf r}'|^2}\frac{e^{-\zeta |\tilde {\mathbf r}|}}{|\tilde {\mathbf r}|}\frac{e^{-\zeta' |\tilde {\mathbf r}'-\tilde {\mathbf r}|}}{|\tilde {\mathbf r}'-\tilde {\mathbf r}|}d\tilde {\mathbf r}d\tilde {\mathbf r}'.
\end{align*}

In order to calculate $\tilde {\mathbf r}$ integral, we introduce the ellipsoidal coordinates. Let $\mathbf q,\mathbf q'\in \mathbb R^3$ be points such that $\mathbf q\neq \mathbf q'$ and set $D=|\mathbf q-\mathbf q'|$, $\tilde r_{\mathbf q}=|\tilde {\mathbf r}-\mathbf q|,\ \tilde r_{\mathbf q'}:=|\tilde {\mathbf r}-\mathbf q'|$. When we choose the direction of $\mathbf q-\mathbf q'$ as the direction of the third axis in $\mathbb R^3$ and $\frac{\mathbf q+\mathbf q'}{2}$ as the origin, ellipsoidal coordinates $(\xi,\eta,\varphi)$ of $\tilde {\mathbf r}$ with foci $\mathbf q,\mathbf q'$ is defined by $\xi=\frac{\tilde r_{\mathbf q}+\tilde r_{\mathbf q'}}{D}$, $\eta=\frac{\tilde r_{\mathbf q}-\tilde r_{\mathbf q'}}{D}$, $\varphi=\mathrm{arccos}(\tilde x/\sqrt{\tilde x^2+\tilde y^2})$, where $\tilde{\mathbf r}=(\tilde x,\tilde y,\tilde z)$. Then the integration of a function $f(\tilde{\mathbf r})$ is written as
\begin{align*}
&\int_{\mathbb R^3}f(\tilde {\mathbf r})d\tilde {\mathbf r}=\frac{D^3}{8}\int_1^{\infty}\left\{\int_{-1}^1\left\{\int_0^{2\pi}(\xi^2-\eta^2)f(\xi,\eta,\varphi)d\varphi\right\}d\eta\right\}d\xi.
\end{align*}
Thus setting $\mathbf q=0,\ \mathbf q'=\tilde {\mathbf r}'$ we obtain
\begin{equation}\label{myeq2.0.0.0.2}
\begin{split}
\int_{\mathbb R^3}\frac{e^{-\zeta |\tilde {\mathbf r}|}}{|\tilde {\mathbf r}|}\frac{e^{-\zeta' |\tilde {\mathbf r}'-\tilde {\mathbf r}|}}{|\tilde {\mathbf r}'-\tilde {\mathbf r}|}d\tilde {\mathbf r}&=\pi|\tilde {\mathbf r}'|\int_{-1}^1\int_1^{\infty}e^{-\frac{\zeta+\zeta'}{2}|\tilde {\mathbf r}'|\xi-\frac{\zeta-\zeta'}{2}|\tilde {\mathbf r}'|\eta}  d \eta d\xi\\
&=\frac{4\pi}{|\tilde {\mathbf r}'|(\zeta^2-\zeta'^2)}(e^{-\zeta'|\tilde {\mathbf r}'|}-e^{-\zeta|\tilde {\mathbf r}'|})\\
&=\frac{4\pi}{\zeta+\zeta'}\int^1_0e^{-(\zeta|\tilde {\mathbf r}'|+(\zeta'-\zeta)|\tilde {\mathbf r}'|t)}dt.
\end{split}
\end{equation}
Thus by the change of variables to the polar coordinates we have
\begin{align*}
\int_{\mathbb R^3}\int_{\mathbb R^3}\frac{1}{|\tilde {\mathbf r}'|^2}\frac{e^{-\zeta |\tilde {\mathbf r}|}}{|\tilde {\mathbf r}|}\frac{e^{-\zeta' |\tilde {\mathbf r}'-\tilde {\mathbf r}|}}{|\tilde {\mathbf r}'-\tilde {\mathbf r}|}d\tilde {\mathbf r}d\tilde {\mathbf r}'&=\frac{4\pi}{\zeta+\zeta'}\int^1_0\int_{\mathbb R^3}\frac{1}{|\tilde {\mathbf r}'|^2}e^{-(\zeta|\tilde {\mathbf r}'|+(\zeta'-\zeta)|\tilde {\mathbf r}'|t)}dtd\tilde {\mathbf r}'\\
&=\frac{16\pi^2}{\zeta+\zeta'}\int^1_0\int_0^{\infty}e^{-(\zeta \tilde r'+(\zeta'-\zeta)\tilde r't)}dtd\tilde r'\\
&=16\pi^2\frac{\log \zeta-\log \zeta'}{\zeta^2-\zeta'^2},
\end{align*}
which completes the proof.

For large $n$ and $n'$ the calculation of the right-hand side of Eq. \eqref{myeq2.0.0.0.1} is not so easy. We are going to derive one efficient method now. Since Eq. \eqref{myeq2.0.0.0.1} is symmetric with respect to $\zeta$ and $\zeta'$, we have only to calculate $\frac{\partial^{n}}{\partial \zeta^{n}}\frac{\partial^{n'}}{\partial \zeta'^{n'}}\frac{\log \zeta}{\zeta^2-\zeta'^2}$. Using binomial coefficients we have
\begin{align*}
\frac{\partial^{n}}{\partial \zeta^{n}}\frac{\partial^{n'}}{\partial \zeta'^{n'}}\frac{\log \zeta}{\zeta^2-\zeta'^2}&=\log \zeta\frac{\partial^{n}}{\partial \zeta^{n}}\frac{\partial^{n'}}{\partial \zeta'^{n'}}\frac{1}{\zeta^2-\zeta'^2}\\
&\quad+\sum_{\nu=0}^{n-1}\binom{n}{\nu}(-1)^{n-\nu-1}(n-\nu-1)!\zeta^{-n+\nu}\frac{\partial^{\nu}}{\partial \zeta^{\nu}}\frac{\partial^{n'}}{\partial \zeta'^{n'}}\frac{1}{\zeta^2-\zeta'^2}.\\
\end{align*}
In order to calculate $\frac{\partial^{\nu}}{\partial \zeta^{\nu}}\frac{\partial^{n'}}{\partial \zeta'^{n'}}\frac{1}{\zeta^2-\zeta'^2}$ we need the following formula for differential operators \cite{NLR} which can be confirmed easily by induction with respect to $n$:
\begin{equation}\label{myeq2.7}
\left(\frac{\partial}{\partial\zeta}\right)^n=\sum_{k=\left[\frac{n+1}{2}\right]}^n\zeta^{2k-n}\beta_k^n\left(\frac{1}{\zeta}\frac{\partial}{\partial\zeta}\right)^k,
\end{equation}
where $\beta_k^n=\frac{2^{k-n}n!}{(n-k)!(2k-n)!}$ and $[t]$ is the greatest integer less than or equal to $t$. Using Eq. \eqref{myeq2.7} we can calculate as
\begin{align*}
\frac{\partial^{\nu}}{\partial \zeta^{\nu}}\frac{\partial^{n'}}{\partial \zeta'^{n'}}\frac{1}{\zeta^2-\zeta'^2}&=\sum_{k=[\frac{\nu+1}{2}]}^{\nu}(-1)^k2^k\zeta^{2k-\nu}\beta^{\nu}_k\sum_{k'=\left[\frac{n'+1}{2}\right]}^{n'}2^{k'}\zeta^{2k'-n'}\beta^{n'}_{k'}\\
&\quad\times(k+k')!\frac{1}{(\zeta^2-\zeta'^2)^{k+k'+1}}.
\end{align*}

\subsection{Method 2: reduction to one-dimensional integrals on a bounded interval}\label{sec2.2}
Even for $l=m=0$ the expression \eqref{myeq2.0.0.0.1} can not be used when $\zeta=\zeta'$, because the denominator is zero and it is difficult to determine the limit as $\zeta\to\zeta'$ in particular for large $n$ and $n'$. However, if we allow existence of one-dimensional integrals of analytic functions on a bounded interval, an expression for arbitrary $l,m$ and $\zeta,\zeta'$ can be derived.

Let us denote the Fourier transform of $f$ by $\mathcal Ff$:
$$\mathcal Ff(\mathbf k)=\int_{\mathbb R^3}e^{-i\mathbf k \cdot \mathbf r}f(\mathbf r)d\mathbf r.$$
We regard $\int_{\mathbb R^3}\frac{1}{|\mathbf r-\mathbf r'|^2}\chi_{lm}^n(\mathbf r,\zeta)d\mathbf r$ as a convolution $\frac{1}{|\mathbf r|^2}*\chi_{lm}^n$ whose Fourier transform is given by $\mathcal F(\frac{1}{|\mathbf r|^2}*\chi_{lm}^n)=(\mathcal F\frac{1}{|\mathbf r|^2})(\mathcal F\chi_{lm}^n)$. Since we have \cite{Sc}
\begin{equation}\label{myeq2.0.0.1}
\left(\mathcal F\frac{1}{|\mathbf r|^2}\right)(\mathbf k)=\frac{2\pi^2}{|\mathbf k|},
\end{equation}
by Perseval's formula we can rewrite the integral as
\begin{equation}\label{myeq2.1.0.1}
[\chi_{lm}^n|\chi_{l'm'}^{n'}]=2^{-2}\pi^{-1}\int_{\mathbb R^3}|\mathbf k|^{-1}\overline{\mathcal F\chi_{lm}^n}(\mathbf k)\mathcal F\chi_{l'm'}^{n'}(\mathbf k)d\mathbf k.
\end{equation}

Now we need an expression of $\mathcal F\chi_{lm}^n$. We shall first calculate $\mathcal F(e^{-r})(\mathbf k)$. In the polar coordinates the Fourier transform is written as
$$
\mathcal F(e^{- r})(\mathbf k)=2\pi\int_0^{\infty}\int_0^{\pi}e^{-irk \cos\theta}e^{-r}r^2\sin \theta drd\theta,
$$
where $k=|\mathbf k|$ and we choose the direction of $\mathbf k$ as the direction of the axis of the polar coordinates on which $\theta=0$. Integration with respect to $r$ can be performed by integration by parts, and we obtain
\begin{align*}
\mathcal F(e^{-r})(\mathbf k)&=2\pi\int_0^{\pi}\frac{2\sin\theta}{(ik\cos\theta+1)^3}d\theta=\frac{8\pi}{(1+k^2)^2}.
\end{align*}
By the change of coordinates $\mathbf r\to\zeta^{-1}\mathbf r$ we have
$$\mathcal F(e^{-\zeta r})(\mathbf k)=\zeta^{-3}\mathcal F(e^{-r})(\zeta^{-1}\mathbf k)=\frac{8\pi\zeta}{(\zeta^2+k^2)^2}.$$
Noting that $r^{-1}e^{-\zeta r}=\int_{\zeta}^{\infty}e^{-t r}dt$ and changing the order of integration we obtain
\begin{equation}\label{myeq2.1.0.2}
\begin{split}
\mathcal F(\chi^0_{00}(\mathbf r,\zeta))(\mathbf k)&=\mathcal F(r^{-1}e^{-\zeta r})(\mathbf k)=\int_{\zeta}^{\infty}\mathcal F(e^{-t r})(\mathbf k)dt\\
&=\int_{\zeta}^{\infty}\frac{8\pi t}{(t^2+k^2)^2}dt=\frac{4\pi}{\zeta^2+k^2}.
\end{split}
\end{equation}
Here we note that 
\begin{equation}\label{myeq2.1}
\chi_{lm}^n=\left(-\frac{\partial}{\partial \zeta}\right)^n\chi_{lm}^0.
\end{equation}
By Eq. \eqref{myeq2.1} we have only to consider $\chi_{lm}^0$. Using the formula for the Fourier transforms of functions multiplied by variables, we can see that the Fourier transform of $\chi_{lm}^0$ is written as
\begin{equation}\label{myeq2.1.1}
\mathcal F(\chi_{lm}^0)(\mathbf k)=Z_l^m(i\nabla_{\mathbf k})\mathcal F(r^{-1}e^{-\zeta r})=i^lZ_l^m(\nabla_{\mathbf k})\mathcal F(r^{-1}e^{-\zeta r}),
\end{equation}
where $\nabla_{\mathbf k}=\left(\frac{\partial}{\partial k_x},\frac{\partial}{\partial k_y},\frac{\partial}{\partial k_z}\right)$.

Now we need to calculate $Z_l^m(\nabla_{\mathbf k})\mathcal F(r^{-1}e^{-\zeta r})$. For this purpose we use the following Hobson's theorem \cite{Ho, Wen2}. Let $f(x,y,z)$ be a homogeneous polynomial of degree $l\in\mathbb N$ in the variables $x,y,z$ and $F\in C^{\infty}(\mathbb R)$. Then we have
\begin{equation}\label{myeq2.1.2}
f(\nabla)F(r)=\sum_{\nu=0}^{\left[\frac{l+1}{2}\right]}\frac{1}{2^\nu \nu !}\left[\left(\frac{1}{r}\frac{d}{dr}\right)^{l-\nu}F(r)\right]\nabla^{2\nu}f(x,y,z).
\end{equation}
If $f$ is a solution to the Laplace equation $\nabla^2 f=0$, only the power $\nabla^{2\nu}$ with $\nu=0$ produces a nonzero result:
\begin{equation}\label{myeq2.2}
f(\nabla)F(r)=\left[\left(\frac{1}{r}\frac{d}{dr}\right)^{l}F(r)\right]f(x,y,z).
\end{equation}
We note here that
\begin{equation}\label{myeq2.3}
\frac{1}{r}\frac{d}{dr}\frac{1}{(s+r^2)^k}=-\frac{2k}{(s+r^2)^{k+1}}.
\end{equation}
Since $Z_l^m$ satisfies the Laplace equation,  using Eqs. \eqref{myeq2.1.0.2}, \eqref{myeq2.2} and \eqref{myeq2.3} we obtain
\begin{equation}\label{myeq2.4}
Z_l^m(\nabla_{\mathbf k})\mathcal F(r^{-1}e^{-\zeta r})=(-1)^l2^{l+2}l!\pi\frac{Z_l^m(\mathbf k)}{(\zeta^2+k^2)^{l+1}}.
\end{equation}
It follows from Eqs. \eqref{myeq2.1}, \eqref{myeq2.1.1} and \eqref{myeq2.4} that
\begin{equation}\label{myeq2.4.1}
\mathcal F(\chi_{lm}^n)=(-i)^l2^{l+2}l!\pi \left(-\frac{\partial}{\partial \zeta}\right)^n\frac{Z_l^m(\mathbf k)}{(\zeta^2+k^2)^{l+1}}.
\end{equation}

Equations \eqref{myeq2.1.0.1} and \eqref{myeq2.4.1} yield
\begin{align*}
[\chi_{lm}^n|\chi_{l'm'}^{n'}]&=i^l(-i)^{l'}2^{l+l'+2}l!l'!\pi\\
&\quad\times\left(-\frac{\partial}{\partial \zeta}\right)^n\left(-\frac{\partial}{\partial \zeta'}\right)^{n'}\int_{\mathbb R^3}\frac{\overline{Z_l^m}(\mathbf k)Z_{l'}^{m'}(\mathbf k)}{k(\zeta^2+k^2)^{l+1}(\zeta'^2+k^2)^{l'+1}}d\mathbf k.
\end{align*}
Using Eq. \eqref{myeq2.0.1} we can see that
\begin{equation}\label{myeq2.5}
\begin{split}
[\chi_{lm}^n|\chi_{l'm'}^{n'}]&=2\delta_{ll'}\delta_{mm'}\alpha_{lm}\\
&\quad\times\left(-\frac{\partial}{\partial \zeta}\right)^n\left(-\frac{\partial}{\partial \zeta'}\right)^{n'}\int_0^{\infty}\frac{k^{2l+1}}{(\zeta^2+k^2)^{l+1}(\zeta'^2+k^2)^{l+1}}dk,
\end{split}
\end{equation}
where
$$\alpha_{lm}=\frac{2^{2l+3}(l!)^2(l+|m|)!\pi^2}{(2l+1)(l-|m|)!}.$$
By the change of the variables $k^2=t$ and $\frac{\zeta'^2}{\zeta'^2+t}=u$ we can see that
\begin{equation}\label{myeq2.6}
\begin{split}
&2\int_0^{\infty}\frac{k^{2l+1}}{(\zeta^2+k^2)^{l+1}(\zeta'^2+k^2)^{l+1}}dk=\int_0^1\frac{u^l(1-u)^l}{(\zeta^2u+\zeta'^2(1-u))^{l+1}}du.
\end{split}
\end{equation}
Combining Eqs. \eqref{myeq2.5}, \eqref{myeq2.6}, \eqref{myeq2.7} and \eqref{myeq2.3} with $r$ replaced by $\zeta$ and $\zeta'$, we finally obtain
\begin{equation}\label{myeq2.8}
\begin{split}
[\chi_{lm}^n|\chi_{l'm'}^{n'}]=(-1)^{n+n'}\delta_{ll'}\delta_{mm'}\alpha_{lm}\sum_{p=\left[\frac{n+1}{2}\right]}^n\zeta^{2p-n}\beta^n_p\sum_{q=\left[\frac{n'+1}{2}\right]}^{n'}\zeta'^{2q-n'}\beta^{n'}_{q}\gamma^{l}_{pq}I^l_{pq}.
\end{split}
\end{equation}
where
$$\gamma^l_{pq}=(-2)^{p+q}\frac{(l+p+q)!}{l!},$$
and
$$I_{pq}^l=\int_0^1\frac{u^{l+p}(1-u)^{l+q}}{(\zeta^2u+\zeta'^2(1-u))^{l+p+q+1}}du.$$

\section{Fundamental two-center integrals}
In this section we consider the following fundamental two-center integrals for $\chi_{lm}^n$ and $\chi_{l'm'}^{n'}$ centered at $\mathbf R_A,\mathbf R_B\in \mathbb R^3,\  \mathbf R_A\neq  \mathbf R_B$:
$$\int_{\mathbb R^3}\int_{\mathbb R^3}\frac{1}{|\mathbf r-\mathbf r'|^2}\overline{\chi_{lm}^n}(\mathbf r_A)\chi_{l'm'}^{n'}(\mathbf r'_B)d\mathbf  rd\mathbf r'.$$
By a translational change of variables we can rewrite the integral as
$$\int_{\mathbb R^3}\int_{\mathbb R^3}\frac{1}{|\mathbf r-\mathbf r'+\mathbf R_A-\mathbf R_B|^2}\overline{\chi_{lm}^n}(\mathbf r)\chi_{l'm'}^{n'}(\mathbf r')d\mathbf rd\mathbf r'.$$
If we choose the direction of $\mathbf R_B-\mathbf R_A$ as the direction of the axis of  the polar coordinates of $\chi_{lm}^n$ and $\chi_{l'm'}^{n'}$, the integral depends only on the distance $R=|\mathbf R_B-\mathbf R_A|$. Thus let us denote the integral by $[\chi_{lm}^n|\chi_{l'm'}^{n'}]_R$, i.e.
\begin{equation}\label{myeq4.1}
\begin{split}
[\chi_{lm}^n|\chi_{l'm'}^{n'}]_R&=\int_{\mathbb R^3}\int_{\mathbb R^3}\frac{1}{|\mathbf r-\mathbf r'-\mathbf R|^2}\overline{\chi_{lm}^n}(\mathbf r)\chi_{l'm'}^{n'}(\mathbf r')d\mathbf rd\mathbf r'\\
&=\int_{\mathbb R^3}\int_{\mathbb R^3}\frac{1}{|\mathbf r-\mathbf r'|^2}\overline{\chi_{lm}^n}(\mathbf r+\mathbf R)\chi_{l'm'}^{n'}(\mathbf r')d\mathbf rd\mathbf r',
\end{split}
\end{equation}
where $\mathbf R=\mathbf R_B-\mathbf R_A$.
Here we note
$$[\chi_{lm}^n|\chi_{l'm'}^{n'}]_R=0,\ m\neq m'.$$
Although this is proved by Parseval's formula and the form Eq. \eqref{myeq2.4.1} of the Fourier transform of $\chi_{lm}^n$, we shall give an elementary proof here. Recall that $\chi_{l'm'}^{n'}$ depends on the angular coordinates $(\theta',\varphi')$ of $\mathbf r'$ through the factor $Y_{l'm'}(\theta',\varphi')$. In the $\mathbf r'$-integral in Eq. \eqref{myeq4.1} we can choose the direction of $\mathbf r$ as a new axis of the polar coordinates of $\mathbf r'$. Then the new polar coordinates of $\mathbf r'$ can be written as $(r',\gamma,\psi)$, where $\gamma$ is the angle between $\mathbf r'$ and $\mathbf r$. Then $Y_{l'm'}(\theta',\varphi')$ can be expanded by the spherical harmonics $Y_{l'\tilde m}(\gamma,\psi)$ with respect to the new coordinates as, \cite{CO2}
$$Y_{l'm'}(\theta',\varphi')=\sum_{m=-l'}^{l'}C_{\tilde m}Y_{l'\tilde m}(\gamma,\psi).$$
Setting $\gamma = 0$ and noting $Y_{l'\tilde m}(0,\psi)=0,\ \tilde m \neq 0$, $Y_{l'0}(0,\psi)=\left (\frac{2l'+1}{4\pi}\right)^{1/2}$, we can see that $C_{0}=\left (\frac{4\pi}{2l'+1}\right)^{1/2}Y_{l'm'}(\theta,\varphi)$, where $(\theta,\varphi)$ is the angular coordinates of $\mathbf r$. Using $\lvert \mathbf r-\mathbf r'\rvert^2 = r^2-2rr'\cos \gamma +r'^2$ we can calculate the $\mathbf r'$-integral as
\begin{equation}\label{myeq4.1.1}
\int_{\mathbb R^3}\frac{1}{|\mathbf r-\mathbf r'|^2}\chi_{l'm'}^{n'}(\mathbf r')d\mathbf r'=f(r)Y_{l'm'}(\theta, \varphi),
\end{equation}
where
\begin{align*}
f(r) =&\frac{4\pi}{2l'+1}\left(\frac{(l'+|m'|)!}{(l'-|m'|)!}\right)^{1/2}\\
&\times\int_{0}^{\infty}\int_0^{\pi}\int_0^{2\pi}\frac{r'^{l'+n'-1}e^{-\zeta' r'}Y_{l'0}(\gamma,\psi)}{r^2-2rr'\cos \gamma +r'^2}dr'd\gamma d\psi.
\end{align*}
Here we used that $\psi$-integral of $Y_{l'\tilde m}(\gamma,\psi),\ \tilde m\neq 0$ vanishes. The result $[\chi_{lm}^n|\chi_{l'm'}^{n'}]_R=0,\ m\neq m'$ can be seen performing $\varphi$-integral in Eq. \eqref{myeq4.1} with the help of Eq. \eqref{myeq4.1.1}.
Therefore, hereafter we consider $[\chi_{lm}^n|\chi_{l'm}^{n'}]_R$ only.

For expressions of $[\chi_{lm}^n|\chi_{l'm}^{n'}]_R$ one needs to deal with products of spherical harmonics. For products of $Y_{lm}$ we have the following formula
\begin{equation}\label{myeq3.4}
Y_{lm}Y_{l'm'}=\sum_{\tilde l=l_{\min}}^{l_{\max}}G_{\tilde l}^{l m l' m'}Y_{\tilde l\ m+m'},
\end{equation}
where $G_{\tilde l}^{l m l' m'}$ is called Gaunt coefficient for which analytic expressions \cite{BL, CO} and an efficient method of computation by recurrence formulas \cite{WS2} are known. The summation limits are given by
\begin{equation}\label{myeq3.4.1}
\begin{split}
l_{\max}&=l+l',\\
l_{\min}&=\begin{cases}
\mu_{\min},\ &\mathrm{if}\ l_{\max}+\mu_{\min}\ \mathrm{is\ even},\\
\mu_{\min}+1,\ &\mathrm{if}\ l_{\max}+\mu_{\min}\ \mathrm{is\ odd},
\end{cases}\\
\mu_{\min}&=\max\{|l-l'|,|m+m'|\}.
\end{split}
\end{equation}
The formula corresponding to Eq. \eqref{myeq3.4} for $Z_l^m$ is
\begin{equation}\label{myeq3.5}
Z_l^mZ_{l'}^{m'}=\sum_{\tilde l=l_{\min}}^{l_{\max}}D^{l m l' m'}_{\tilde l}G_{\tilde l}^{l m l' m'}r^{2\Delta l}Z_{\tilde l}^{m+m'},
\end{equation}
where $\Delta l=(l+l'-\tilde l)/2$ and
\begin{align*}
D^{l m l' m'}_{\tilde l}=\left(\frac{4\pi(2\tilde l+1)(l+|m|)!(l'+|m'|)!(\tilde l-|m+m'|)!}{(2l+1)(2l'+1)(l-|m|)!(l'-|m'|)!(\tilde l+|m+m'|)!}\right)^{1/2}.
\end{align*}
Here note that considering the parity of functions one has $G_{\tilde l}^{l m l' m'}\neq 0$ only if $l+l'-\tilde l$ is even, and thus $\Delta l$ is a natural number.

\subsection{Method 1: analytic expression}
In order to obtain an analytic expression of $[\chi_{lm}^n|\chi_{l'm}^{n'}]_R$ we apply the shift-operator approach \cite{NLR2}. We define a differential operator $\Omega^n_{lm}(\nabla_A,\zeta)$ by
$$\Omega^n_{lm}(\nabla_A,\zeta)=Z_l^m(\nabla_A)\left(-\frac{\partial}{\partial\zeta}\right)^n\left(-\frac{1}{\zeta}\frac{\partial}{\partial\zeta}\right)^l,$$
where $\nabla_A=(\frac{\partial}{\partial X_A},\frac{\partial}{\partial Y_A},\frac{\partial}{\partial Z_A})$ with the Cartesian coordinates $(X_A,Y_A,Z_A)$ of $\mathbf R_A$. Then by Hobson's theorem Eq. \eqref{myeq2.2} and
$$\left(-\frac{1}{r}\frac{\partial}{\partial r}\right)\left(-\frac{1}{\zeta}\frac{\partial}{\partial\zeta}\right)(r^{-1}e^{-\zeta r})=r^{-1}e^{-\zeta r},$$
we can see that
$$\chi^n_{lm}(\mathbf r_A,\zeta)=\Omega^n_{lm}(\nabla_A,\zeta)(r_A^{-1}e^{-\zeta r_A}),$$
where $r_A=|\mathbf r_A|=|\mathbf r-\mathbf R_A|$.
Thus we can rewrite $[\chi_{lm}^n|\chi_{l'm'}^{n'}]_R$ as
\begin{equation}\label{myeq3.0.1}
\begin{split}
[\chi_{lm}^n|\chi_{l'm}^{n'}]_R&=\overline{\Omega^n_{lm}}(\nabla_A,\zeta)\Omega^{n'}_{l'm}(\nabla_B,\zeta')\int_{\mathbb R^3}\int_{\mathbb R^3}\frac{1}{|\mathbf r-\mathbf r'|^2}\frac{e^{-\zeta |\mathbf r-\mathbf R_A|}}{|\mathbf r-\mathbf R_A|}\frac{e^{-\zeta' |\mathbf r'-\mathbf R_B|}}{|\mathbf r'-\mathbf R_B|}d\mathbf rd\mathbf r'\\
&=\overline{\Omega^n_{lm}}(\nabla_A,\zeta)\Omega^{n'}_{l'm}(\nabla_B,\zeta')[\chi_{00}^0|\chi_{00}^{0}]_R,
\end{split}
\end{equation}
where
$$\overline{\Omega^n_{lm}}(\nabla_A,\zeta)=\overline{Z_l^m}(\nabla_A)\left(-\frac{\partial}{\partial\zeta}\right)^n\left(-\frac{1}{\zeta}\frac{\partial}{\partial\zeta}\right)^l.$$

For $[\chi_{00}^0|\chi_{00}^{0}]_R$ we have the following expression whose proof is given in the appendix:
\begin{equation}\label{myeq3.0.2}
[\chi_{00}^0|\chi_{00}^{0}]_R=\frac{8\pi^2}{\zeta^2-\zeta'^2}\left(\frac{g(\zeta'R)}{\zeta'R}-\frac{g(\zeta R)}{\zeta R}\right).
\end{equation}
Here
$$g(t)=e^{-t}\mathrm{Ei}(t)-e^{t}\mathrm{Ei}(-t),$$
where $\mathrm{Ei}(t)$ is the exponential integral defined by
$$\mathrm{Ei}(t)=-\mathrm{p.v.}\int_{-t}^{\infty}\frac{e^{-s}}{s}ds.$$

With the help of Eq. \eqref{myeq3.5} and the equation $\overline{Z_l^m}=(-1)^mZ_l^{-m}$ the differential operator $\overline{Z_l^m}(\nabla_A)Z_{l'}^{m}(\nabla_B)$ in Eq. \eqref{myeq3.0.1} applied to a function $f(R)$ of $R$ is written as
\begin{align*}
\overline{Z_l^m}(\nabla_A)Z_{l'}^{m}(\nabla_B)f(R)=(-1)^{l+m}\sum_{\tilde l=l_{\min}}^{l_{\max}}D^{l -m l' m}_{\tilde l}G_{\tilde l}^{l -m l' m}\nabla_B^{2\Delta l}Z_{\tilde l}^{0}(\nabla_B)f(R),
\end{align*}
Using Hobson's theorem Eq. \eqref{myeq2.1.2} and 
$$\nabla^2_{\mathbf R}(R^{2\Delta l}Z_{\tilde l}^0(\mathbf R))=2\Delta l(2\Delta l+2\tilde l+1)R^{2\Delta l-2}Z_{\tilde l}^0(\mathbf R),$$
The last expression is rewritten as
\begin{equation}\label{myeq3.0.3}
\begin{split}
\overline{Z_l^m}(\nabla_A)Z_{l'}^{m}(\nabla_B)f(R)&=(-1)^{l+m}\sum_{\tilde l=l_{\min}}^{l_{\max}}D^{l -m l' m}_{\tilde l}G_{\tilde l}^{l -m l' m}\sum_{p=0}^{\Delta l}E^{\tilde l \Delta l}_pR^{2\Delta l-2p}Z_{\tilde l}^{0}(\mathbf R)\\
&\quad \times\left(\frac{1}{R}\frac{\partial}{\partial R}\right)^{l+l'-p}f(R),
\end{split}
\end{equation}
where
$$E^{\tilde l \Delta l}_p=\frac{2^{p}\Delta l!\Gamma(\Delta l+\tilde l+3/2)}{p!(\Delta l-p)!\Gamma(\Delta l+\tilde l-p+3/2)}.$$

Combining Eqs. \eqref{myeq3.0.1}-\eqref{myeq3.0.3} one obtains
\begin{equation}\label{myeq3.0.3.1}
\begin{split}
[\chi_{lm}^n|\chi_{l'm}^{n'}]_R&=(-1)^{n+n'+l'+m+1}8\pi^2\sum_{\tilde l=l_{\min}}^{l_{\max}}D^{l -m l' m}_{\tilde l}G_{\tilde l}^{l -m l' m}\sum_{p=0}^{\Delta l}E^{\tilde l \Delta l}_pR^{2\Delta l-2p}Z_{\tilde l}^{0}(\mathbf R)\\
&\quad\times(U^{n l n' l'}_{l+l'-p}(R,\zeta,\zeta')+U^{n' l' n l}_{l+l'-p}(R,\zeta',\zeta)),
\end{split}
\end{equation}
where
\begin{align*}
U^{n l n' l'}_{q}(R,\zeta,\zeta')&=\left(\frac{\partial}{\partial\zeta}\right)^n\left(\frac{1}{\zeta}\frac{\partial}{\partial\zeta}\right)^l\left(\frac{\partial}{\partial\zeta'}\right)^{n'}\left(\frac{1}{\zeta'}\frac{\partial}{\partial\zeta'}\right)^{l'}\left(\frac{1}{R}\frac{\partial}{\partial R}\right)^{q}\frac{g(\zeta R)}{(\zeta^2-\zeta'^2)\zeta R}\\
&=\sum_{\mu=0}^n\binom{n}{\mu}\sum_{\nu=0}^{l}\binom{l}{\nu}v_{n' l'}^{n-\mu\ l-\nu}(\zeta,\zeta')w_q^{\mu\nu}(R,\zeta),
\end{align*}
with
\begin{align*}
v_{n_2 l_2}^{n_1 l_1}(\zeta,\zeta')&=\left(\frac{\partial}{\partial\zeta}\right)^{n_1}\left(\frac{1}{\zeta}\frac{\partial}{\partial\zeta}\right)^{l_1}\left(\frac{\partial}{\partial\zeta'}\right)^{n_2}\left(\frac{1}{\zeta'}\frac{\partial}{\partial\zeta'}\right)^{l_2}\frac{1}{\zeta^2-\zeta'^2},\\
w_q^{\mu\nu}(R,\zeta)&=\left(\frac{\partial}{\partial\zeta}\right)^{\mu}\left(\frac{1}{\zeta}\frac{\partial}{\partial\zeta}\right)^{\nu}\left(\frac{1}{R}\frac{\partial}{\partial R}\right)^{q}\frac{g(\zeta R)}{\zeta R}.
\end{align*}
With the help of Eqs. \eqref{myeq2.7} and \eqref{myeq2.3} we obtain
\begin{align*}
v_{n_2 l_2}^{n_1 l_1}(\zeta,\zeta')&=(-1)^{l_1}\sum_{\lambda_1=\left[\frac{n_1+1}{2}\right]}^{n_1}(-1)^{\lambda_1}2^{\lambda_1+l_1}\beta_{\lambda_1}^{n_1}\zeta^{2\lambda_1-n_1}\\
&\quad\times\sum_{\lambda_2=\left[\frac{n_2+1}{2}\right]}^{n_2}2^{\lambda_2+l_2}\beta_{\lambda_2}^{n_2}\zeta^{2\lambda_2-n_2}\frac{(\lambda_1+l_1+\lambda_2+l_2)!}{(\zeta^2-\zeta'^2)^{\lambda_1+l_1+\lambda_2+l_2+1}}.
\end{align*}
Using Eq. \eqref{myeq2.7} one finds
$$w_q^{\mu\nu}(R,\zeta)=\sum_{\sigma=\left[\frac{\mu+1}{2}\right]}^{\mu}\beta^{\mu}_{\sigma}\zeta^{2\sigma-\mu}\tilde w^{\nu+\sigma}_q(R,\zeta),$$
where
$$\tilde w^{s}_q(R,\zeta)=\left(\frac{1}{\zeta}\frac{\partial}{\partial\zeta}\right)^{s}\left(\frac{1}{R}\frac{\partial}{\partial R}\right)^{q}\frac{g(\zeta R)}{\zeta R}.$$

Since $\tilde w_q^{s}(R,\zeta)$ is symmetric with respect to the exchange of the pairs $(\zeta, s)$ and $(R, q)$, it remains to derive an expression for $\tilde w_q^{s}(R,\zeta)$ with $s\geq q$.
Here we need the following formula for operators:
\begin{equation}\label{myeq3.0.4}
\left(\frac{1}{R}\frac{\partial}{\partial R}\right)^{q}\frac{1}{R}=\sum_{j=0}^qC_j^qR^{2j-2q-1}\left(\frac{1}{R}\frac{\partial}{\partial R}\right)^{j},
\end{equation}
with
$$C_j^q=\frac{2^jq!\prod_{i=0}^q(2j-2i+1)}{(q-j)!(2j+1)!},$$
and
\begin{equation}\label{myeq3.0.5}
\left(\frac{1}{\zeta}\frac{\partial}{\partial\zeta}\right)^{\tau}=\sum_{\kappa=1}^{\tau}(-1)^{\tau-\kappa}B_{\kappa}^{\tau}\zeta^{\kappa-2\tau}\left(\frac{\partial}{\partial\zeta}\right)^{\kappa},
\end{equation}
with
$$B_{\kappa}^{\tau}:=\frac{(2\tau-\kappa-1)!}{2^{\tau-\kappa}(\tau-\kappa)!(\kappa-1)!},$$
which can easily be confirmed by induction with respect to $q$ and $\tau$ respectively. Note that $\frac{1}{R}$ in Eq. \eqref{myeq3.0.4} is a multiplication operator, and that the left hand side does not mean application of $\left(\frac{1}{R}\frac{\partial}{\partial R}\right)^{q}$ to $\frac{1}{R}$. Combining Eq. \eqref{myeq3.0.4}, the equation
$$\frac{1}{\zeta}\frac{\partial}{\partial\zeta}\frac{1}{R}\frac{\partial}{\partial R}\frac{g^{(M)}(\zeta R)}{\zeta}=\frac{g^{(M+2)}(\zeta R)}{\zeta},$$
and Eq. \eqref{myeq3.0.5} one has
\begin{equation}\label{myeq3.0.5.1}
\begin{split}
\tilde w^{s}_q(R,\zeta)&=\sum_{j=0}^qC_j^qR^{2j-2q-1}\left(\frac{1}{\zeta}\frac{\partial}{\partial\zeta}\right)^{s-j+1}\frac{g^{(2j-1)}(\zeta R)}{R}\\
&=\sum_{j=0}^qC_j^q\sum_{\kappa=1}^{s-j+1}(-1)^{s-j+1-\kappa}B_{\kappa}^{s-j+1}\zeta^{\kappa-2(s-j+1)}R^{2j-2q+\kappa-2} g^{(2j+\kappa-1)}(\zeta R).
\end{split}
\end{equation}
The derivatives of $g$ in the last expression is expressed by direct calculations as
\begin{align*}
g^{(2M)}(t)&=-\sum_{i=1}^M\frac{2(2i-2)!}{t^{2i-1}}+g(t),\\
g^{(2M+1)}(t)&=\sum_{j=1}^M\frac{2(2i-1)!}{t^{2i}}-e^{t}\mathrm{Ei}(-t)-e^{-t}\mathrm{Ei}(t).
\end{align*}

\subsection{Method 2: reduction to one-dimensional integrals}
Using Parseval's formula, Eqs. \eqref{myeq2.7}, \eqref{myeq2.4.1}, \eqref{myeq3.5} and $\mathcal F(\chi_{lm}^n(x_A))=e^{-i\mathbf R_A\cdot \mathbf k}\mathcal F(\chi_{lm}^n(x))$ we obtain
$$[\chi_{lm}^n|\chi_{l'm}^{n'}]_R=A^{ll'}_m\sum_{\tilde l=l_{\min}}^{l_{\max}}(2\tilde l+1)^{1/2}G^{l -m l' m}_{\tilde l}M^{nn'}_{l l'\tilde l},$$
where
\begin{align*}
A^{ll'}_m&=(-1)^{l'+m}i^{l+l'}2^{l+l'+3}l!l'!\pi\left(\frac{\pi(l+|m|)!(l'+|m|)!}{(2l+1)(2l'+1)(l-|m|)!(l'-|m|)!}\right)^{1/2},
\end{align*}
and
\begin{align*}
M^{nn'}_{l l'\tilde l}&=(-1)^{n+n'}\sum_{p=\left[\frac{n+1}{2}\right]}^n\beta^n_p\zeta^{2p-n}(-2)^p\frac{(l+p)!}{l!}\\
&\quad\times\sum_{p'=\left[\frac{n'+1}{2}\right]}^{n'}\beta^{n'}_{p'}\zeta'^{2p'-n'}(-2)^{p'}\frac{(l'+p')!}{l'!}L^{l+p\ l'+p'}_{\tilde l \, \Delta l},\\
\end{align*}
with
$$L^{l+p\ l'+p'}_{\tilde l\, \Delta l}=\int_{\mathbb R^3}e^{-i\mathbf R\cdot\mathbf k}\frac{k^{2\Delta l-1}Z^0_{\tilde l}(\mathbf k)}{(\zeta^2+k^2)^{l+p+1}(\zeta'^2+k^2)^{l'+p'+1}}d\mathbf k.$$
This integral is again a Fourier transform. For the evaluation of this Fourier transform we use the Rayleigh expansion of a plane wave in terms of spherical Bessel functions and spherical harmonics
$$e^{-i\mathbf R\cdot\mathbf k}=4\pi\sum_{\hat l=0}^{\infty}\sum_{m=-\hat l}^{\hat l}(-i)^{\hat l}j_{\hat l}(R k)Y_{\hat lm}(\theta_{\mathbf R},\varphi_{\mathbf R})Y_{\hat lm}^*(\theta_{\mathbf k},\varphi_{\mathbf k}),$$
where $j_{\hat l}$ is the spherical Bessel function. \cite{Wei, WS3} With the help of this expansion we obtain
\begin{equation}\label{myeq4.2}
\begin{split}
L^{l+p\ l'+p'}_{\tilde l\, \Delta l}=&(2\pi)^{3/2}(-i)^{\tilde l}R^{l+l'+2p+2p'+2}\\
&\times\int_0^{\infty}\frac{k^{l+l'+1/2}J_{\tilde l+1/2}(k)}{((\zeta R)^2+k^2)^{l+p+1}((\zeta' R)^2+k^2)^{l'+p'+1}}dk,
\end{split}
\end{equation}
where $J_{\tilde l +1/2}$ is the Bessel function.

\section{Method for three and four-center integrals}\label{subsec6.1}
Three and four-center integrals are reduced to one or two-center integrals with quantitative error bounds. We consider general integral $[\psi_1(\mathbf r_A)\psi_2(\mathbf r_B)|\psi_3(\mathbf r'_C)\psi_4\newline(\mathbf r'_D)]$, where each $\psi_i$ has the form of $\chi_{lm}^n$. In $[\psi_1(\mathbf r_A)\psi_2(\mathbf r_B)|\psi_3(\mathbf r'_C)\psi_4(\mathbf r'_D)]$ we expand $\psi_2(\mathbf r_B)$ and $\psi_4(\mathbf r'_D)$ by STOs $\varphi_j$ centered at $\mathbf R_A$ and $\mathbf R_C$ respectively:
\begin{align*}
\psi_2(\mathbf r_B) &= \sum_{j=1}^{\infty} c_j\varphi_j(\mathbf r_A),\\
\psi_4(\mathbf r'_D) &= \sum_{k=1}^{\infty} \tilde c_k\varphi_k(\mathbf r'_C).
\end{align*}
As $\varphi_j$ we can choose a complete orthonormal system \cite{RL2}. Each $\varphi_j$ can be written as a linear combination of $\chi_{lm}^n$. Thus with the help of the Gaunt coefficients each $[\psi_1(\mathbf r_A)\varphi_j(\mathbf r_A)|\psi_3(\mathbf r'_C)\varphi_k(\mathbf r'_C)]$ can be written as a finite sum of fundamental one or two-center integrals. Since in the practical calculation we need to truncate the expansions up to a finite sum $\Phi_J(\mathbf r_A)=\sum_{j=1}^{J} c_j\varphi_j(\mathbf r_A)$ and $\tilde \Phi_K(\mathbf r'_C)=\sum_{k=1}^{K}\tilde c_k\varphi_k(\mathbf r'_C)$, we have to estimate the error by the truncation written as follows:
\begin{equation}\label{myeq6.1}
\begin{split}
&[\psi_1(\mathbf r_A)\psi_2(\mathbf r_B)|\psi_3(\mathbf r'_C)\psi_4(\mathbf r'_D)]-[\psi_1(\mathbf r_A)\Phi_J(\mathbf r_A)|\psi_3(\mathbf r'_C)\tilde \Phi_K(\mathbf r'_C)]\\
&\quad =[\psi_1(\mathbf r_A)(\psi_2(\mathbf r_B)-\Phi_J(\mathbf r_A))|\psi_3(\mathbf r'_C)\psi_4(\mathbf r'_D)]\\
&\qquad + [\psi_1(\mathbf r_A)\Phi_J(\mathbf r_A)|\psi_3(\mathbf r'_C)(\psi_4(\mathbf r'_D)-\tilde\Phi_K(\mathbf r'_C))].
\end{split}
\end{equation}

For the estimate we use the $L^{\infty}$-norm defined by $\lVert \psi\rVert_{L^{\infty}}=\sup_{\mathbf r\in \mathbb R^3}\lvert \psi(\mathbf r)\rvert$,
in addition to the usual $L^2$-norm.
The norm $\lVert\chi_{lm}^n\rVert_{L^{\infty}}$ of $\chi_{lm}^n$ can be evaluated easily. Since $\varphi_j$ is an orthonormal system, the $L^2$-norms of $\Phi_J$ and $\psi_2(\mathbf r_B)-\Phi_J(\mathbf r_A)$ are also evaluated as
\begin{align*}
\lVert\Phi_J\rVert^2&=\sum_{j=1}^J \lvert c_j\rvert^2,\\
\lVert\psi_2(\mathbf r_B)-\Phi_J(\mathbf r_A)\rVert^2&=\lVert\psi_2\rVert^2-\sum_{j=1}^J \lvert c_j\rvert^2.
\end{align*}
By the Schwarz inequality the first term in the right-hand side of Eq. \eqref{myeq6.1} is estimated as
\begin{align*}
&\left \lvert[\psi_1(\mathbf r_A)(\psi_2(\mathbf r_B)-\Phi_J(\mathbf r_A))|\psi_3(\mathbf r'_C)\psi_4(\mathbf r'_D)]\right \rvert\\
&\quad\leq \lVert  \psi_1^*(\mathbf r_A)(\psi_2(\mathbf r_B)-\Phi_J(\mathbf r_A))\rVert\left\lVert \frac{1}{r^2}*(\psi_3^*(\mathbf r_C)\psi_4(\mathbf r_D))\right\rVert\\
&\quad\leq\lVert  \psi_1\rVert_{L^{\infty}}\lVert\psi_2(\mathbf r_B)-\Phi_J(\mathbf r_A)\rVert\left\lVert \frac{1}{r^2}*(\psi_3^*(\mathbf r_C)\psi_4(\mathbf r_D))\right\rVert.
\end{align*}
Using Eq. \eqref{myeq2.0.0.1} and the Hardy inequality the last factor is estimated as
\begin{align*}
\left\lVert \frac{1}{r^2}*(\psi_3^*(\mathbf r_C)\psi_4(\mathbf r_D))\right\rVert&=\left\lVert \frac{1}{r^2}*(\psi_3^*(\mathbf r)\psi_4(\mathbf r_{CD}))\right\rVert\\
&=2^{-1/2}\pi^{1/2}\left\lVert \frac{1}{k}\mathcal F(\psi_3^*(\mathbf r)\psi_4(\mathbf r_{CD}))(\mathbf k)\right\rVert\\
&\leq (2\pi)^{1/2}\left\lVert \nabla_{\mathbf k}\mathcal F(\psi_3^*(\mathbf r)\psi_4(\mathbf r_{CD}))(\mathbf k)\right\rVert\\
&=4\pi^{2}\left\lVert r\psi_3^*(\mathbf r)\psi_4(\mathbf r_{CD})\right\rVert\\
&\leq4\pi^{2}\lVert r\psi_3(\mathbf r)\rVert_{L^{\infty}}\lVert\psi_4\rVert,
\end{align*}
where $\mathbf r_{CD}=\mathbf r-\mathbf R_D+\mathbf R_C$. Thus we obtain
\begin{align*}
&\left\lvert[\psi_1(\mathbf r_A)(\psi_2(\mathbf r_B)-\Phi_J(\mathbf r_A))|\psi_3(\mathbf r'_C)\psi_4(\mathbf r'_D)]\right\rvert\\
&\quad \leq4\pi^{2}\lVert  \psi_1\rVert_{L^{\infty}}\lVert\psi_2(\mathbf r_B)-\Phi_J(\mathbf r_A)\rVert\lVert r\psi_3(\mathbf r)\rVert_{L^{\infty}}\lVert\psi_4\rVert.
\end{align*}

In the same way the second term in the right-hand side of Eq. \eqref{myeq6.1} is estimated as
\begin{align*}
&\left\lvert [\psi_1(\mathbf r_A)\Phi_J(\mathbf r_A)|\psi_3(\mathbf r'_C)(\psi_4(\mathbf r'_D)-\tilde \Phi_K(\mathbf r'_C))]\right\rvert\\
 &\quad\leq4\pi^2\lVert  \psi_1\rVert_{L^{\infty}}\lVert\Phi_J\rVert\lVert r\psi_3(\mathbf r)\rVert_{L^{\infty}}\lVert\psi_4(\mathbf r_D)-\tilde\Phi_K(\mathbf r_C)\rVert.
\end{align*}
Thus we have the error estimate for the truncation as
\begin{align*}
&\left\lvert[\psi_1(\mathbf r_A)\psi_2(\mathbf r_B)|\psi_3(\mathbf r'_C)\psi_4(\mathbf r'_D)]-[\psi_1(\mathbf r_A)\Phi_J(\mathbf r_A)|\psi_3(\mathbf r'_C)\tilde \Phi_K(\mathbf r'_C)]\right\rvert\\
&\quad\leq4\pi^2\lVert  \psi_1\rVert_{L^{\infty}}\lVert r\psi_3(\mathbf r)\rVert_{L^{\infty}}\\
&\qquad\times\left (\lVert\psi_2(\mathbf r_B)-\Phi_J(\mathbf r_A)\rVert\lVert\psi_4\rVert+\lVert\Phi_J\rVert\lVert\psi_4(\mathbf r_D)-\tilde \Phi_K(\mathbf r_C)\rVert\right).
\end{align*}
In practical calculations we need to choose $J$ and $K$ large enough so that the last factor will be small enough.

\section{fundamental hybrid two-center integrals}\label{subsec6.2}
Fundamental hybrid two-center integrals are defined by
\begin{align*}
&[\chi^{n_1}_{l_1m_1}|\chi^{n_2}_{l_2m_2}\chi^{n_3}_{l_3m_3}]_R\\
&\quad=\int_{\mathbb R^3}\int_{\mathbb R^3}\overline{\chi^{n_1}_{l_1m_1}}(\mathbf r_A,\zeta_1)\frac{1}{|\mathbf r-\mathbf r'|^2}\chi^{n_2}_{l_2m_2}(\mathbf r'_A,\zeta_2)\chi^{n_3}_{l_3m_3}(\mathbf r'_B,\zeta_3)d\mathbf r d\mathbf r'\\
&\quad=\int_{\mathbb R^3}\int_{\mathbb R^3}\overline{\chi^{n_1}_{l_1m_1}}(\mathbf r,\zeta_1)\frac{1}{|\mathbf r-\mathbf r'|^2}\chi^{n_2}_{l_2m_2}(\mathbf r',\zeta_2)\chi^{n_3}_{l_3m_3}(\mathbf r'_{AB},\zeta_3)d\mathbf r d\mathbf r',
\end{align*}
where $\mathbf R_A\neq\mathbf R_B$ and $r'_{AB}=\mathbf r'-\mathbf R_B+\mathbf R_A$. The integral Eq. \eqref{myeq1.1} with $\mathbf R_A = \mathbf R_B=\mathbf R_C\neq\mathbf R_D$ is reduced to integrals of this form using the Gaunt coefficient.
We can evaluate this integral expanding $\chi^{n_3}_{l_3m_3}(\mathbf r'_{AB},\zeta_3)$ by STOs centered at $0$ by the method in Section \ref{subsec6.1} and evaluating the resulting one-center integrals by the method in Section \ref{thirdsec}. For the expansion of $\chi^{n_3}_{l_3m_3}(\mathbf r'_{AB},\zeta_3)$ we use the following formula \cite{RL3}
\begin{equation}\label{myeq6.1.1}
P_L^M(\cos \Theta)r_{AB}^{N-1}e^{-r_{AB}}=\sum_{k=M}^{\infty}\sum_{p=0}^{\infty}C_{kp}^{NLM}\omega^p_{kM}(\mathbf r,\mu),
\end{equation}
for $N,L,M\in\mathbb N,\ L\geq M$, $\mu>0$, where $r_{AB}=|\mathbf r_{AB}|$ and
$$\omega^p_{kM}(\mathbf r, \mu)=P_k^{M}(\cos \theta)(2\mu r)^{k}e^{-\mu r}L^{2k+2}_p(2\mu r).$$
Here $L^{2k+2}_p(2\mu r)$ is the associated Laguerre polynomial \cite{GR}. The coefficients $C_{kp}^{NLM}$ can be calculated by recurrence relations depending on $\mu$ and $R=|\mathbf R_B-\mathbf R_A|$. 
Since the functions $\omega^p_{kM}(\mathbf r, \mu),\ k=M,M+1,\dots,\ p=0,1,\dots$ form a complete orthogonal system, we can apply the arguments in Section \ref{subsec6.1}. 
The formula Eq. \eqref{myeq6.1.1} and the expression of the Laguerre polynomial yield the expansion
\begin{equation}\label{myeq6.1.2}
\begin{split}
\chi^{n_3}_{l_3m_3}(\mathbf r_{AB},\zeta_3)&=i^{m_3+|m_3|}e^{im_3\varphi}\zeta_3^{-n_3-l_3+1}\sum_{k=|m_3|}^{\infty}\sum_{p=0}^{\infty}C_{kp}^{(n_3+l_3)l_3|m_3|}\omega^p_{k|m_3|}(\mathbf r,\zeta_3)\\
&=\zeta_3^{-n_3-l_3+1}\sum_{k=M}^{\infty}\sum_{p=0}^{\infty}\sum_{q=0}^pT_{kpq}^{n_3l_3m_3}\chi_{km_3}^{q+1}(\mathbf r,\zeta_3),
\end{split}
\end{equation}
where $T^{n_3l_3m_3}_{kpq}=C^{(n_3+l_3)l_3|m_3|}_{kp}(-1)^{q}\binom{2k+p+2}{p-q}\frac{1}{q!}(2\zeta_3)^{k+q}$, and $C^{(n_3+l_3)l_3|m_3|}_{kp}$ depends on $\zeta_3R$.
Using this expansion we obtain
$$[\chi^{n_1}_{l_1m_1}|\chi^{n_2}_{l_2m_2}\chi^{n_3}_{l_3m_3}]_R=\zeta_3^{-n_3-l_3+1}\sum_{k=|m_3|}^{\infty}\sum_{p=0}^{\infty}\sum_{q=0}^pT^{n_3l_3m_3}_{kpq}[\chi^{n_1}_{l_1m_1}|\chi^{n_2}_{l_2m_2}\chi^{q+1}_{km_3}],$$
where
\begin{equation*}
[\chi^{n_1}_{l_1m_1}|\chi^{n_2}_{l_2m_2}\chi^{q+1}_{km_3}]=\int_{\mathbb R^3}\int_{\mathbb R^3}\overline{\chi^{n_1}_{l_1m_1}}(\mathbf r,\zeta_1)\frac{1}{|\mathbf r-\mathbf r'|^2}\chi^{n_2}_{l_2m_2}(\mathbf r',\zeta_2)\chi^{q+1}_{km_3}(\mathbf r',\zeta_3)d\mathbf r d\mathbf r'.
\end{equation*}
Using Eq. \eqref{myeq3.5} we can see that
\begin{equation}\label{myeq6.2}
\begin{split}
&[\chi^{n_1}_{l_1m_1}|\chi^{n_2}_{l_2m_2}\chi^{q+1}_{km_3}]\\
&\quad=\delta_{m_1(m_2+m_3)}D_{l_1}^{l_2m_2 k m_3}G_{l_1}^{l_2m_2 k m_3}[\chi^{n_1}_{l_1m_1}|\chi^{n_2+q+l_2+k-l_1}_{l_1 m_1}(\zeta_2+\zeta_3)],
\end{split}
\end{equation}
for $l_2+l_3\geq l_1\geq l_{\min}$ and it vanishes in the other cases. Here $[\chi^{n_1}_{l_1m_1}|\chi^{n_2+q+l_2+k-l_1}_{l_1 m_1}\newline(\zeta_2+\zeta_3)]$ is the fundamental one-center integral with the index $\zeta_2+\zeta_3$ of the second STO, and $l_{\min}$ is the natural number defined by \eqref{myeq3.4.1} with $l, l', m, m'$ replaced by $l_2, k, m_2, m_3$. From Eq. \eqref{myeq6.2} we can see that $[\chi^{n_1}_{l_1m_1}|\chi^{n_2}_{l_2m_2}\chi^{n_3}_{l_3m_3}]_R=0$ unless $m_1=m_2+m_3$. In practical calculations we truncate the expansion of $\chi^{n_3}_{l_3m_3}(\mathbf r_{AB})$ in Eq. \eqref{myeq6.1.2} up to finite terms. Let us denote the finite sum by $\Phi_J(\mathbf r)$ as in Section \ref{subsec6.1}, that is, if we use the terms up to $k=k_{\max}$ and $p=p_{\max}$,
$$\Phi_J(\mathbf r)=i^{m_3+|m_3|}e^{im\varphi}\zeta_3^{-n_3-l_3+1}\sum_{k=|m_3|}^{k_{\max}}\sum_{p=0}^{p_{\max}}C_{kp}^{(n_3+l_3)l_3|m_3|}\omega^p_{k|m_3|}(\mathbf r,\zeta_3).$$
Following the arguments in Section \ref{subsec6.1} we have the error bound of the truncation
\begin{equation}\label{myeq6.3}
\begin{split}
&|[\chi^{n_1}_{l_1m_1}|\chi^{n_2}_{l_2m_2}\chi^{n_3}_{l_3m_3}]_R-[\chi^{n_1}_{l_1m_1}|\chi^{n_2}_{l_2m_2}\Phi_J]|\\
&\quad\leq4\pi^2\lVert r\chi^{n_1}_{l_1m_1}(\mathbf r)\rVert\lVert\chi^{n_2}_{l_2m_2}\rVert_{L^{\infty}}\lVert \chi^{n_3}_{l_3m_3}(\mathbf r_{AB})-\Phi_J(\mathbf r)\rVert\\
&\quad\leq 8\pi^2\sqrt{\frac{\pi(2n_1+2l_1+2)!(l_1+|m_1|)!}{(2\zeta_1)^{2n_1+2l_1+3}(2l_1+1)(l_1-|m_1|)!}}\left(\frac{n_2+l_2-1}{\zeta_2}\right)^{n_2+l_2-1}\frac{(l_2+|m_2|)!}{l_2!}\\
&\qquad\times e^{-n_2-l_2+1}\lVert \chi^{n_3}_{l_3m_3}(\mathbf r_{AB})-\Phi_J(\mathbf r)\rVert,
\end{split}
\end{equation}
where
$$[\chi^{n_1}_{l_1m_1}|\chi^{n_2}_{l_2m_2}\Phi_J]=\int_{\mathbb R^3}\int_{\mathbb R^3}\overline{\chi^{n_1}_{l_1m_1}}(\mathbf r,\zeta_1)\frac{1}{|\mathbf r-\mathbf r'|^2}\chi^{n_2}_{l_2m_2}(\mathbf r',\zeta_2)\Phi_J(\mathbf r')d\mathbf r d\mathbf r'.$$
Here for the estimate of $\lVert\chi^{n_2}_{l_2m_2}\rVert_{L^{\infty}}$ we used the following formula \cite{GR2},
$$P_l^m(z)=\frac{(-1)^m(l+m)!}{l!\pi}\int_0^{\pi}\left(z+\sqrt{z^2-1}\cos\varphi\right)^l\cos m\varphi d\varphi,$$
(note that the coefficient of our definition of $P_l^m(z)$ and that in the reference are different by the factor $(-1)^m$)
and estimated the associated Legendre function as
$$|P_l^m(\cos \theta)|\leq \frac{(l+m)!}{l!\pi}\int_0^{\pi}d\varphi=\frac{(l+m)!}{l!}.$$

\section{Numerical results}
The fundamental one-center integrals were evaluated for $\zeta=1$, $\zeta'=0.5$ by the analytic expression and the expression by one-dimensional integrals. As for the one-dimensional integrals the integral $I_{pq}^l$ in Eq. \eqref{myeq2.8} was evaluated approximating the integrand by the Chebyshev interpolation with typical order $1000$ and integrating the polynomial. Calculations with high orders by this method are extremely easy because the zeros of the Chebyshev polynomials have easy analytic expressions. This method is slightly different from the Chebyshev-Gauss quadrature and gives much better results for simple analytic integrands than the Chebyshev-Gauss quadrature. This method would be an ordinary way, but specific name for the method could not be found. Accurate significant figures were obtained by determining invariant figures by varying the order of the Chebyshev interpolation. All calculations were performed with double precision. 
For $l=m=0$ the number of the accurate figures $F_{ae}$ of the value by the analytic expression are also shown in Table \ref{mytab2}.  The number $F_{ae}$ was determined comparing the value obtained by using the expression Eq. \eqref{myeq2.0.0.0.1} and the value by the one dimensional integrals obtained above as a reliable reference. Examples are presented in Table \ref{mytab2}.

\begin{table}[htb]
\begin{center}
\caption{The accurate significant figures of $[\chi_{lm}^n|\chi_{lm}^{n'}]$}\label{mytab2}
\setlength{\tabcolsep}{5pt}
\begin{tabular}{ccccrc} \hline
$n$ & $ n'$ & $l$ & $m$ & $[\chi_{lm}^n|\chi_{lm}^{n'}]$ & $F_{ae}$\\ \hline
2 & 3 & 0 & 0 & $1.56939270526650(4)$ & 14\\
2 & 3 & 5 & 4 & $3.425716931848(16)$\\
2 & 3 & 10 & 9 & $1.0469905487775(39)$\\
4 & 4 & 0 & 0 & $1.953591848090(6)$ & 13\\
4 & 4 & 5 & 4 & $5.8161756391883(19)$\\
4 & 4 & 10 & 9 & $6.7706640231478(42)$\\
6 & 5 & 0 & 0 & $6.77033700568(8)$ & 11\\
6 & 5 & 5 & 4 & $1.5712472039294(23)$\\
6 & 5 & 10 & 9 & $5.9442801255419(46)$\\
8 & 8 & 0 & 0 & $8.8795833287(13)$ & 9\\
8 & 8 & 5 & 4 & $2.22546915631(29)$\\
8 & 8 & 10 & 9 & $4.332795650516(53)$\\
11 & 10 & 0 & 0 & $5.5789551(19)$ & 7\\
11 & 10 & 5 & 4 & $7.1529791758(35)$\\
11 & 10 & 10 & 9 & $5.881306549(60)$\\
14 & 14 & 0 & 0 & $2.5509(28)$ & 4\\
14 & 14 & 5 & 4 & $3.91588207(45)$\\
14 & 14 & 10 & 9 & $1.642355950(71)$\\
\hline
\end{tabular}
\begin{tablenotes}
\item \footnotesize{The notation $(\nu)$ signifies $\times10^{\nu}$.}
\end{tablenotes}
\end{center}
\end{table}

Examining each step of the calculation, it was observed that the loss of accuracy in the expression by one-dimensional integrals was due to the cancellation of significant digits in the summations in Eq. \eqref{myeq2.8}. The cancellation was less and the result was more accurate often  for large $l$ than for small $l$.

The fundamental two-center integrals were also evaluated by the two expressions for $R=4$, $\zeta=1$, $\zeta'=0.5$.
As for the one-dimensional integrals the integral $L^{l+p\ l'+p'}_{\tilde l\, \Delta l}$ in Eq. \eqref{myeq4.2} was evaluated using the Chebyshev interpolation with typical order $1000$ as in the case of one-center integral. Typically integration on the interval $[0,100]$ was enough, because that on $[100,\infty)$ was relatively very small and negligible owing to the decay of the integrands. All calculations were performed with double precision. 
The accurate significant figures of $[\chi_{lm}^n|\chi_{l'm}^{n'}]_R$ were obtained by determining invariant figures varying the order of the Chebyshev interpolation and the interval of the integration of $L^{l+p\ l'+p'}_{\tilde l\, \Delta l}$. The number of the accurate figures $F_{ae}$ of the value by the analytic expression was determined comparing the value obtained by using the expression Eq. \eqref{myeq3.0.3.1} and the value obtained above as a reliable reference. Examples are presented in Table \ref{mytab1}.
\begin{table}[htb]
\begin{center}
\caption{The accurate significant figures of $[\chi_{lm}^n|\chi_{l'm}^{n'}]_R$}\label{mytab1}
\setlength{\tabcolsep}{5pt}
\begin{tabular}{cccccrc} \hline
$n$ & $l$ & $ n'$ & $ l'$ & $m$ & $[\chi_{lm}^n|\chi_{l'm}^{n'}]_R$ & $F_{ae}$\\ \hline
3 & 2 & 3 & 2 & 1 & $2.2243751772625(7)$ & 10\\
3 & 2 & 3 & 3 & 1 & $-2.7566722179287(8)$ & 10\\
2 & 4 & 4 & 5 & 4 & $-1.3610327905104(16)$ & 6\\
2 & 4 & 4 & 6 & 4 & $2.0420467016732(17)$ & 4\\
2 & 5 & 2 & 6 & 4 & $-5.4090782928132(16)$ & 3\\
2 & 6 & 2 & 6 & 4 & $4.986742283667(17)$ & 1\\
2 & 7 & 2 & 6 & 5 & $2.73141199999476(20)$ & 1\\
2 & 7 & 2 & 7 & 5 & $8.0955544928731(21)$ & 0\\
5 & 9 &  5 & 10 & 3 & $-1.0629407232265(33)$ & 0\\
5 & 10 &  5 & 10 & 3 & $9.83626880416(33)$ & 0\\
10 & 5 & 10 & 4 & 2 & $3.8326037195(29)$ & 0\\
10 & 5 & 10 & 5 & 2 & $6.9193761122(31)$ & 0\\
\hline
\end{tabular}
\begin{tablenotes}
\item \footnotesize{The notation $(\nu)$ signifies $\times10^{\nu}$.}
\end{tablenotes}
\end{center}
\end{table}

In contrast to the high accuracy of the method by one-dimensional integrals, the accuracy of the analytic expression deteriorates rapidly as $l$, $l'$, $n$ and $n'$ increase, and the results are completely meaningless for the parameters greater than moderate values. It was observed that in the calculation of $\tilde w_q^{s}$ in Eq. \eqref{myeq3.0.5.1} enormous cancellations of significant digits happen.

Finally the fundamental hybrid two-center integrals were evaluated by the method in Subsection \ref{subsec6.2}. For the evaluation of one-center integrals in the right hand side of Eq. \eqref{myeq6.2} we used the expression by one-dimensional integrals as above. Here recall that $[\chi_{l_1m_1}^{n_1}|\chi_{l_2m_2}^{n_2}\chi_{l_3m_3}^{n_3}]_R=0$ unless $m_1=m_2+m_3$. Examples for $\zeta_1=1.0, \zeta_2 = 0.5, \zeta_3=1.0,\ R=0.5$ are presented in Table \ref{mytab3}. Terms in Eq. \eqref{myeq6.1.2} corresponding to $k\leq15$ and $p\leq 15$ were used for the calculation. The error bounds of the errors by this truncation given after $\pm$ in the Table \ref{mytab3} were calculated from Eq. \eqref{myeq6.3}. It was confirmed that the number of significant figures of the finite sum in the expansion of $[\chi_{l_1m_1}^{n_1}|\chi_{l_2m_2}^{n_2}\chi_{l_3m_3}^{n_3}]_R$ which were determined changing the order of the one-dimensional integrals is greater than the number of meaningful figures from the viewpoint of the error bound by Eq. \eqref{myeq6.3}.
\begin{table}[htb]
\begin{center}
\caption{The accurate significant figures of $[\chi_{l_1(m_2+m_3)}^{n_1}|\chi_{l_2m_2}^{n_2}\chi_{l_3m_3}^{n_3}]_R$}\label{mytab3}
\setlength{\tabcolsep}{5pt}
\begin{tabular}{ccccccccr} \hline
$n_1$ & $l_1$ & $ n_2$ & $ l_2$ & $m_2$ & $ n_3$ & $ l_3$ & $m_3$ & $[\chi_{l_1(m_2+m_3)}^{n_1}|\chi_{l_2m_2}^{n_2}\chi_{l_3m_3}^{n_3}]_R$\\ \hline
1 & 1 & 1 & 1 & 0 & 3 & 2 & 1 & $2.00918\pm0.00042(3)$\\
1 & 1 & 1 & 1 & 0 & 4 & 2 & 1 & $8.49388\pm0.00039(3)$\\
1 & 2 & 1 & 1 & 1 & 3 & 2 & 1 & $-5.6044\pm0.0084(3)$\\
1 & 2 & 1 & 2 & 1 & 4 & 2 & 1 & $5.77148\pm0.00035(5)$\\
3 & 2 & 1 & 1 & 0 & 4 & 2 & 1 & $-1.8072\pm0.0054(4)$\\
3 & 2 & 1 & 2 & 1 & 4 & 2 & 1 & $1.573420\pm0.000095(7)$\\
\hline
\end{tabular}
\begin{tablenotes}
\item \footnotesize{The notation $(\nu)$ signifies $\times10^{\nu}$.}
\end{tablenotes}
\end{center}
\end{table}

\appendix
\section{}
In this appendix we prove Eq. \eqref{myeq3.0.2}. First the change of the variable $\tilde {\mathbf r}=\mathbf r-\mathbf r'$ in Eq. \eqref{myeq4.1} with $n=l=m=0$ yields
\begin{equation}\label{myeqa.0.1}
[\chi_{00}^0|\chi_{00}^0]_R=\int_{\mathbb R^3}\int_{\mathbb R^3}\frac{1}{|\tilde{\mathbf r}-\mathbf R|^2}\frac{e^{-\zeta|\mathbf r|}}{|\mathbf r|}\frac{e^{-\zeta'|\mathbf r-\tilde{\mathbf r}|}}{|\mathbf r-\tilde{\mathbf r}|}d\mathbf rd\tilde{\mathbf r}
\end{equation}
The $\mathbf r$ integral can be performed in the same way as in Eq. \eqref{myeq2.0.0.0.2} and yields
\begin{equation}\label{myeqa.1}
\begin{split}
\int_{\mathbb R^3}\frac{e^{-\zeta|\mathbf r|}}{|\mathbf r|}\frac{e^{-\zeta'|\mathbf r-\tilde{\mathbf r}|}}{|\mathbf r-\tilde{\mathbf r}|}d\mathbf r=\frac{4\pi}{|\tilde {\mathbf r}|(\zeta^2-\zeta'^2)}(e^{-\zeta'|\tilde {\mathbf r}|}-e^{-\zeta|\tilde {\mathbf r}|}).
\end{split}
\end{equation}

Hence it remains to calculate the $\tilde{\mathbf r}$ integral.
Changing the variable to the ellipsoidal coordinates with foci $0$ and $\mathbf R$ we have
$$\int_{\mathbb R^3}\frac{1}{|\tilde{\mathbf r}-\mathbf R|^2}\frac{e^{-\zeta|\tilde {\mathbf r}|}}{|\tilde{\mathbf r}|}d\tilde{\mathbf r}=2\pi\int_{-1}^1\int_1^{\infty}\frac{e^{-\zeta\frac{R}{2}(\xi+\eta)}}{\xi-\eta}d\xi d\eta.$$
Next we change the variable as $t=\xi+\eta$, $s=\xi-\eta$ and obtain
\begin{equation}\label{myeqa.2}
\begin{split}
\int_{\mathbb R^3}\frac{1}{|\tilde{\mathbf r}-\mathbf R|^2}\frac{e^{-\zeta|\tilde {\mathbf r}|}}{|\tilde{\mathbf r}|}d\tilde{\mathbf r}&=\pi\Bigg(\int_0^2\int_{2-t}^{2+t}\frac{e^{-\zeta \frac{R}{2}t}}{s}dsdt+\int_2^{\infty}\int_{t-2}^{t+2}\frac{e^{-\zeta \frac{R}{2}t}}{s}dsdt\Bigg)\\
&=\pi\Bigg(\int_0^2e^{-\zeta \frac{R}{2}t}(\log(2+t)-\log(2-t))dt\\
&\quad+\int_2^{\infty}e^{-\zeta \frac{R}{2}t}(\log(t+2)-\log(t-2))dt\Bigg).
\end{split}
\end{equation}
The integral for the integrands including the factor $\log(t+2)$ is easily calculated by integration by parts as
\begin{equation}\label{myeqa.3}
\begin{split}
\pi\int_0^{\infty}e^{-\zeta \frac{R}{2}t}\log(t+2)dt&=\frac{2\pi\log2}{\zeta R}+\frac{2\pi}{\zeta R}\int_0^{\infty}\frac{e^{-\zeta \frac{R}{2}t}}{t+2}dt\\
&=\frac{2\pi\log2}{\zeta R}-\frac{2\pi}{\zeta R}e^{\zeta R}\mathrm{Ei}(-\zeta R).
\end{split}
\end{equation}
The integrals for the factors $\log(2-t)$ and $\log(t-2)$ are improper integrals and require attention. They are expressed as a limit and calculated as
\begin{equation}\label{myeqa.4}
\begin{split}
&-\pi\lim_{\epsilon\to +0}\Bigg(\int_0^{2-\epsilon}e^{-\zeta \frac{R}{2}t}\log(2-t)dt+\int_{2+\epsilon}^{\infty}e^{-\zeta \frac{R}{2}t}\log(t-2)dt\Bigg)\\
&\quad=-\pi\lim_{\epsilon\to +0}\Bigg(-\frac{2\log\epsilon}{\zeta R}e^{-\zeta \frac{R}{2}(2-\epsilon)}+\frac{2\log 2}{\zeta R}-\frac{2}{\zeta R}\int_0^{2-\epsilon}\frac{e^{-\zeta \frac{R}{2}t}}{2-t}dt\\
&\qquad+\frac{2\log\epsilon}{\zeta R}e^{-\zeta \frac{R}{2}(2+\epsilon)}+\frac{2}{\zeta R}\int_{2+\epsilon}^{\infty}\frac{e^{-\zeta \frac{R}{2}t}}{t-2}dt\Bigg)\\
&\quad=-\frac{2\pi\log2}{\zeta R}+\frac{2\pi}{\zeta R}e^{-\zeta R}\mathrm{Ei}(\zeta R).
\end{split}
\end{equation}
Combining Eqs. \eqref{myeqa.2}-\eqref{myeqa.4} we obtain
\begin{equation}\label{myeqa.5}
\int_{\mathbb R^3}\frac{1}{|\tilde{\mathbf r}-\mathbf R|^2}\frac{e^{-\zeta|\tilde {\mathbf r}|}}{|\tilde{\mathbf r}|}d\tilde{\mathbf r}=\frac{2\pi}{\zeta R}e^{-\zeta R}\mathrm{Ei}(\zeta R)-\frac{2\pi}{\zeta R}e^{\zeta R}\mathrm{Ei}(-\zeta R).
\end{equation}
Equation \eqref{myeq3.0.2} immediately follows from Eqs. \eqref{myeqa.0.1}, \eqref{myeqa.1} and \eqref{myeqa.5}.

\end{document}